\documentclass[onecolumn,floatfix,superscriptaddress,a4paper,
               showpacs,showkeys,nofootinbib,preprint]{revtex4}

\usepackage{epsfig}
\usepackage{latexsym}
\usepackage{xspace}
\usepackage{hyperref}
\usepackage[latin2]{inputenc}
\usepackage{indentfirst}
\usepackage{enumerate}

\usepackage{amsmath}
\usepackage{amssymb}
\usepackage[english]{babel}
\usepackage{url}

\newcommand{\bs}{\boldsymbol}

\begin{document}

\title{Phase Transition in Interacting Boson System at Finite Temperatures}
\author{D. Anchishkin}
\affiliation{
Bogolyubov Institute for Theoretical Physics, 03680 Kiev, Ukraine}
\affiliation{
Taras Shevchenko National University of Kiev, 03022 Kiev, Ukraine}
\affiliation{
Frankfurt Institute for Advanced Studies, Ruth-Moufang-Strasse 1,
60438 Frankfurt am Main, Germany}
\author{I. Mishustin}
\affiliation{
Frankfurt Institute for Advanced Studies, Ruth-Moufang-Strasse 1,
60438 Frankfurt am Main, Germany}
\affiliation{
Johann Wolfgang Goethe University, D-60438 Frankfurt am Main, Germany}
\author{H. Stoecker}
\affiliation{
Frankfurt Institute for Advanced Studies, Ruth-Moufang-Strasse 1,
60438 Frankfurt am Main, Germany}
\affiliation{
Johann Wolfgang Goethe University, D-60438 Frankfurt am Main, Germany}

\date{\today}



\begin{abstract}
Thermodynamical properties of an interacting boson system at finite
temperatures and zero chemical potential are studied within the framework of the
Skyrme-like mean-field toy model.
It is assumed that the mean field contains both attractive and repulsive terms.
Self-consistency relations between the mean field and thermodynamic
functions are derived.
It is shown that for sufficiently strong attractive interactions this system
develops a first-order phase transition via formation of Bose condensate.
An interesting prediction of the model is that the condensed phase is
characterized by a constant total density of particles.
The thermodynamical characteristics of the system are calculated for the
liquid-gas and condensed phases.
The energy density exhibits a jump at the critical temperature.
\end{abstract}

\maketitle

\section{Introduction}
\label{sec1}
In recent years the properties of hot and dense hadronic matter have
attracted considerable interest. Such matter can be produced in
relativistic nucleus-nucleus collisions which are under
investigations in many laboratories. QCD motivated effective models
and lattice simulations indicate that the chiral symmetry
restoration and the deconfinement phase transition should take place
at high temperatures and particle densities.
In hot medium, the properties of the hadrons are expected to be modified.
In this paper we study specifically the properties of interacting boson systems
in the framework of a toy model using thermodynamically
consistent mean-field approach.
The model aims to estimate the scale and relation of attractive and repulsive
contributions to the potential in order to investigate a possibility of the
Bose-Einstein condensation of interacting bosonic particles.
This problem has been studied previously, starting from the pioneer works of
A.B.~Migdal and coworkers \cite{migdal-1972,migdal-1974,migdal-1978,saperstein-1990}
and later by many authors using different models and methods.
A possible formation of classical pion fields in relativistic nucleus-nucleus
collisions was discussed in
refs.~\cite{anselm-1991,blaizot-1992,bjorken-1992,mishustin-greiner-1993}.
In more recent studies
\cite{son-2001,kogut-2001,toublan-2001,mammarella-2015,carignano-2017},
pionic systems with a finite isospin chemical potential at low
temperatures have been considered.
Interesting new results concerning dense pionic systems have been obtained
recently using lattice methods \cite{brandt-2016,brandt-2017}.

In the present paper we consider interacting boson systems at zero chemical
potential, but high temperatures, where thermally produced particles have
rather high densities.
Simple calculations for noninteracting hadron resonance gas show that the
particle density may reach values $(0.1 - 0.2)$~fm$^{-3}$ at temperatures
$100 - 150$~MeV, which are below the deconfinement phase transition,
see e.g. refs.~\cite{satarov-2009,vovchenko-2017}.
Under such conditions the interaction effects should become important.
To account for the interaction between the bosons we introduce a
phenomenological Skyrme-like mean field $U(n)$, which depends on the
particle density $n$ only.
Then the thermodynamical consistency relations are used to calculate the
particle density, energy density and pressure as functions of temperature.
An important difference of the considered system is that, in contrast to e.g.
the bosonic matter, the number of bosons is not conserved, but is determined by
the minimization of a thermodynamic potential.

The paper is organized as follows.
Section~\ref{sec:mfm} shortly describes the thermodynamic mean-field model which is
used in the presented calculations.
The self-consistency relations between the mean field and thermodynamic functions
are derived.
In Section~\ref{sec:interacting-ps} we introduce a Skyrme-like parametrization
of the mean field and calculate corresponding thermodynamic functions.
In Section~\ref{sec:condensed-phase} we demonstrate the possibility of the
Bose condensation when the attractive interaction is strong enough.
Our conclusions are summarized in Section~\ref{sec:conclusions}.

\section{Thermodynamically consistent mean-field model}
\label{sec:mfm}

First we shortly remind the basics of the thermodynamical mean-field model which
was introduced in ref.~\cite{anch-vovchenko-2014}, see more details in
\cite{anch-1992,anchsu-1995}.

Let us consider the system of interacting particles from general thermodynamic
point of view.
One can describe such a system in terms of the
free energy density $\phi (n,T)$,
which depends on particle density $n$ and temperature $T$.
The free energy density (FED) is related to other thermodynamical quantities
as follows
\footnote{Here and below we adopt the system of units $\hbar = c = 1$, $k_{_B} = 1$}
\begin{eqnarray}
\label{eq:phi-1}
\phi (n,T) &=& \varepsilon(n,T)\, -\, T\, s(n,T)\,,
\\
\phi (n,T) &=& n\, \mu(n,T)\, -\, p(n,T)\,,
\label{eq:phi-2}
\end{eqnarray}
where $\varepsilon(n,T)$ is the energy density and $p(n,T)$ is the pressure.
Two quantities $\mu(n,T)$ (the chemical potential) and  $s(n,T)$ (the entropy density)
are given as partial derivatives with respect to independent variables $(n,\, T)$
\begin{eqnarray}
\mu \ =\ \left( \frac{\partial \phi }{\partial n} \right)_T \,, \qquad
s\ =\ -\, \left( \frac{\partial \phi }{\partial T} \right)_n \,.
\label{eq:phi-3}
\end{eqnarray}

Very generally, for a system of interacting particles the FED can be
written as a sum of free and interacting contributions
\begin{equation}
\phi (n,T)\ =\ \phi_0(n,T)\, +\, \phi_{\rm int}(n,T)\ ,
\label{eq:phi-4}
\end{equation}
where $\phi_0$ is the FED of the noninteracting system.
The chemical potential can also be splitted  into ``free'' and ``interacting''
pieces.
In accordance with eq.~(\ref{eq:phi-3}) we obtain
\begin{equation}
\mu \, =\, \mu_0 + \left( \frac{\partial \phi _{\rm int}}{\partial n} \right) _{T} \ ,
\quad {\rm where} \quad
\mu_0\  \equiv \ \left(\frac{\partial \phi_0}{\partial n}\right)_T \ .
\label{eq:mu-int}
\end{equation}
Further, taking into account eqs. (\ref{eq:phi-2}), (\ref{eq:phi-4}) and
(\ref{eq:mu-int}) one can represent the pressure in the following form
\begin{equation}
p\ =\
n\, \mu(n,T)\, -\, \phi(n,T) \
=\
p_0(n,T) + n\, \left( \frac{\partial \phi _{\rm int}}{\partial n} \right)_T
- \phi _{\rm int}\ ,
\label{eq:pressure-int}
\end{equation}
where
\begin{equation}
p_0(n,T)\ =\ n\, \mu_0(n,T)\, -\, \phi_0(n,T) \ .
\label{eq:pressure0-1}
\end{equation}
One can put this expression in correspondence to
the pressure of the ideal gas $\widetilde{p}_0$
calculated in the grand canonical ensemble for the same values $T$ and $\mu_0$
as they are taken in (\ref{eq:pressure0-1})
\begin{equation}
\widetilde{p}_0(T,\mu_0)\, =\, \frac{g}{3} \int \frac{d^3k}{(2\pi )^3}\,
\frac{{\bs k}^2}{\sqrt{{m^2 + \bs k}^2} }\, f(\bs k;T,\mu_0) \ ,
\label{eq:pressure0-2}
\end{equation}
where $g$ is the degeneracy factor, $f(\bs k;T,\mu_0)$ is the ideal gas
distribution function (the Boltzmann, Fermi-Dirac or Bose-Einstein one),
which depends on temperature and ideal chemical potential $\mu_0$.

Now we introduce the following important notations:
\begin{eqnarray}
\label{eq:mean-field}
U(n,T) & = &
\left[ \frac{\partial \phi_{\rm int}(n,T)}{\partial n} \right]_T \ ,
\\
P^{\rm ex}(n,T) & = &
n \, \left[ \frac{\partial \phi_{\rm int}(n,T)}{\partial n} \right]_T \,
-\, \phi_{\rm int}(n,T)\ .
\label{eq:ex-pressure}
\end{eqnarray}
One can immediately see that these quantities are related as
\begin{equation}
n \frac{\partial U(n,T)}{\partial n}\
=\ \frac{\partial P^{\rm ex}(n,T)}{\partial n} \ .
\label{eq:consistency}
\end{equation}

By subtracting eq.~(\ref{eq:ex-pressure}) from eq.~(\ref{eq:pressure-int}) one
can rewrite the total pressure as
\begin{equation}
p\ =\ p_0(n,T) + P^{\rm ex}(n,T)\ .
\label{eq:pressure-int-2}
\end{equation}
Evidently, if one defines $p_0(n,T)$ as the pressure of the
ideal gas, then, the quantity $P^{\rm ex}(n,T)$ should be regarded as an
{\it excess pressure}, which is due to the interaction between particles.

Next, in our evaluations of the thermodynamic quantities of the interacting
system we would like to use formula (\ref{eq:pressure0-2}) for the pressure
of the ideal gas.
In the canonical ensemble the independent variables are $n$ and $T$,
whereas in the grand canonical ensemble they are $\mu$ and $T$.
Hence, it is necessary to express the free chemical potential $\mu_0$ through
these variables.
One can do this by substituting eq.~(\ref{eq:mean-field}) into eq.~(\ref{eq:mu-int}),
thus obtaining
\begin{equation}
\mu\ =\ \mu_0 \, +\, U(n,T) \ .
\label{eq:mu0-2}
\end{equation}
It is convenient to introduce also the single-particle energy for interacting
particles
\begin{equation}
E(\bs k,n)\ =\ \sqrt{m^2 + {\bs k}^2}\, +\, U(n) \ .
\label{eq:spe}                              
\end{equation}
In the grand canonical ensemble we treat the particle density $n$ as
$n(\mu,T)$, and as a result, the pressure of interacting particles can be
expressed as
\begin{equation}
p(T,\mu )\, =\,
 \frac{g}{3}\, \int \frac{d^3k}{(2\pi)^3}\, \frac{{\bs k}^2}
{\sqrt{m^2 + {\bs k}^2} }\, f(\bs k;T,\mu) + P^{\rm ex}(n,T) \,,
\label{eq:pressure-int-3}
\end{equation}
where
\begin{equation}
f(\bs k;T,\mu)\, =\,  \left\{ \exp{\left[ \frac{E(\bs k,n)-\mu }{T}\right]}
+ a \right\}^{-1}
\label{eq:df}
\end{equation}
with $a=+1$ for fermions, $a=-1$ for bosons and $a=0$ for the Boltzmann
statistics.

In a homogeneous system, where the thermodynamic potential can be expressed as $
\Omega(T,\mu, V) = - p(T,\mu)V$, the particle density
reads $n(T,\mu) = \partial p(T,\mu)/\partial \mu$.
Then, with the help of relations (\ref{eq:consistency}) and
(\ref{eq:pressure-int-3}) we obtain the standard relation
\begin{equation}
n\ =\ g  \int \frac{d^3k}{(2\pi)^3}\, f(\bs k;T,\mu) \,.
\label{eq:n}
\end{equation}
Since the distribution function $f(\bs k;T,\mu)$ is itself a function of $n$,
this expression  is not a formula for particle
density $n$, but an equation to be solved in a selfconsistent way for every point
of the $(T,\mu)$ plane.
The solution will result in the explicit dependence $n=n(T,\mu )$, which in
general differs from the ideal gas expression, $n_0(T,\mu_0)$.
Using $s = \partial p/\partial T$ and the Euler relation,
$\varepsilon + p = Ts + \mu n$, one obtains for the energy density
\begin{eqnarray}
\varepsilon(T,\mu ) & = & g \int \frac{d^3k}{(2\pi )^3}\,
\sqrt{m^2 + {\bs k}^2}\, f(\bs k;T,\mu)\, +\,  n\, U(n,T)\, +\,  n\, \mu\,
-\, P^{\rm ex}(n,T)\,  +
\nonumber \\
&& +\, T\left\{ \left[ \frac{\partial P^{\rm ex}(n,T)}{\partial T}\right]_n\,
-\, n\, \left[ \frac{\partial U(n,T)}{\partial T} \right]_n \right\} \,.
\label{eq:energy-den-int}
\end{eqnarray}

This type of approach is widely used in relativistic mean field models of nuclear
matter
\cite{wal,ser,wald}, where the nucleons interact with a scalar field
$\varphi$ (attraction) and a vector field $V_{\mu }$ (repulsion).
In the homogeneous static system
only the ``time'' component of the vector field $V_0$ survives.
In our notations it is equal to the mean field $U(n,T)$.

\section{Application for interacting boson systems with   
zero chemical potential }
\label{sec:interacting-ps}

\subsection{Skyrme-like parametrization of the mean field }
The formalism described in the previous section has been applied for several
physically interesting systems including the hadron-resonance gas
\cite{anch-vovchenko-2014} and the pionic gas \cite{anch-2016}.
In the present study we extend this approach to the case of a bosonic system
which potentially can undergo Bose condensation.
To illustrate this possibility we formulate the Skyrme-like toy model where
we assume that the interaction between particles is described by the mean field
\begin{equation}
U(n)\, =\, - \, A\,n\, +\, B\, n^2 \,,
\label{eq:mf1}
\end{equation}
where $n$ is the particle density, $A$ and $B$ are the positive model parameters,
which should be specified for each particle species.
A similar approach was used recently in Ref.~\cite{satarov-2017} for describing
Bose condensation in the system of interacting alpha-particles.

As well known, see e.g. ref.~\cite{gasser-1989},
the pion-pion interaction at low energies is rather small, because of their special
nature as the Goldstone bosons for spontaneously broken chiral symmetry.
For the isospin-symmetric pion system, the first coefficient
can be expressed as $A = -\frac{4\pi}{3} (a_0^0 + 5a_0^2)$, where
$a_0^0$ and $a_0^2$ are the $\pi \pi$ $s$-wave scattering lengths for isospin 0 and 2.
Their values are well established both theoretically and experimentally:
$a_0^0 = 0.220\, \frac{1}{m_\pi}$, $a_0^2 = 0.044\, \frac{1}{m_\pi}$,
see e.g. ref.~\cite{conangelo-2001}.
With these values the coefficient $A$ practically vanishes.
Nevertheless, some additional contribution to the attractive mean field at high
temperatures, ($T \ge 150$~MeV), may be provided by other hadrons present in the
system, like $\rho$-mesons\cite{shuryak-1991} or baryon-antibaryon pairs  \cite{Theis}.

By this reason, in our calculations we consider a general case of $A > 0$,
to study a bosonic system with both attractive and repulsive contributions to the mean
field (\ref{eq:mf1}).
To be specific in numerical calculations we will take bosons with
mass $m_{\pi}=140$ MeV and degeneracy factor $g=3$, which we call "pions".
For the repulsive coefficient $B > 0$ we use a fixed value, obtained from
an estimate based on the virial expansion \cite{hansen-2005},
$B = 10 m_\pi b^2$ with $b$ equal to four times the proper volume of a
particle, i.e. $b = 16 \pi r_0^3/3$.
Below we take $b = 0.45$~fm$^3$ that corresponds to a pion
radius $r_0 \approx 0.3$~fm.
The attractive coefficient $A$ is considered as a model parameter.

\begin{figure}
\begin{center}
\centering
\includegraphics[width=0.7\textwidth]{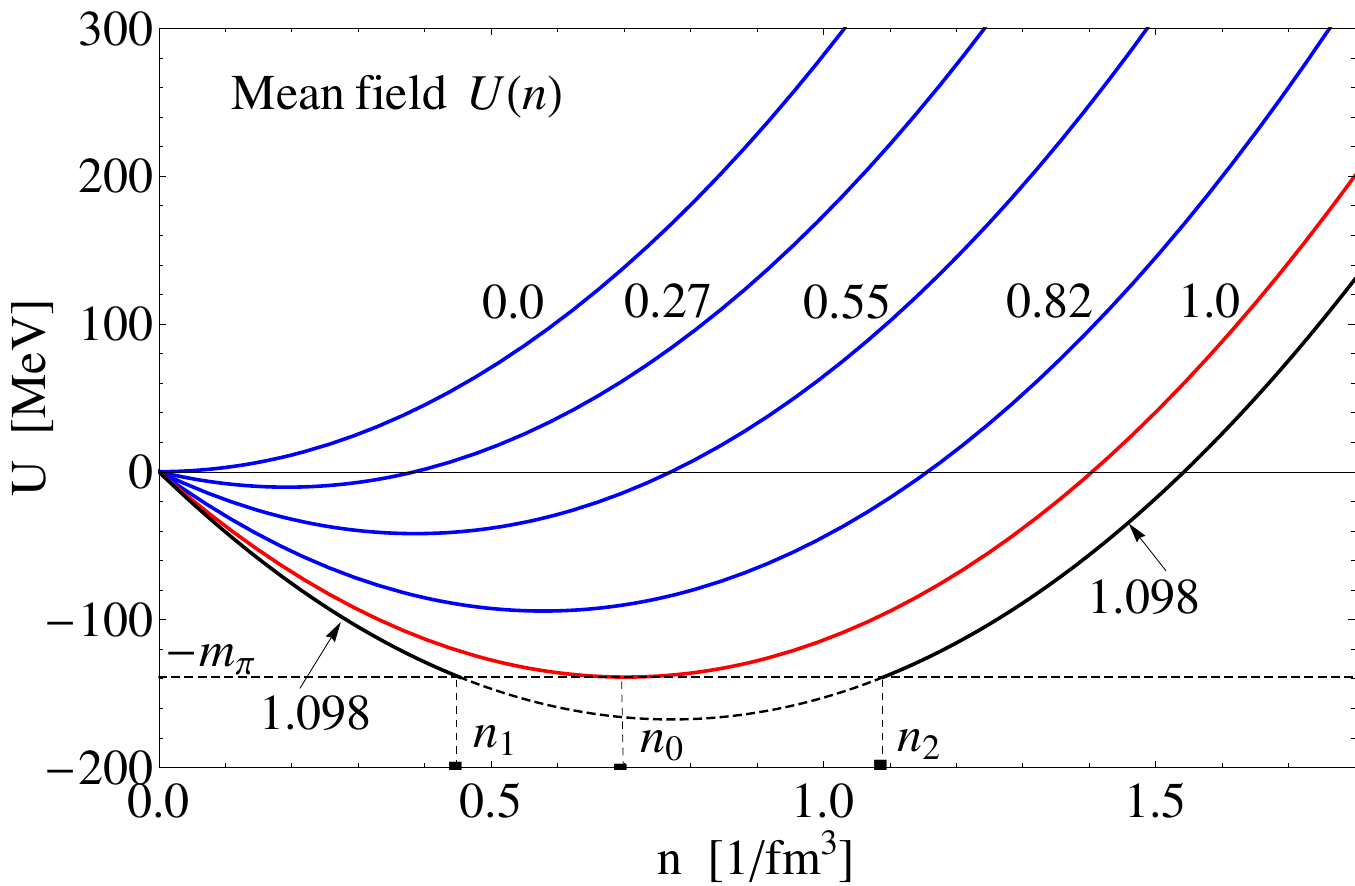}
\centering
\caption{The mean field potential $U$ versus particle density $n$ for an
interacting pion system with $\mu = 0$.
The blue curves are labeled by value of the parameter $\kappa$:
$\kappa = 0$, 
$\kappa = 0.27$, 
$\kappa = 0.55$, 
$\kappa = 0.82$. 
In the case $\kappa = \kappa_{\rm c} = 1.0$ (red solid curve),
the corresponding curve touches the negative energy level $-m_\pi$.
For the case $\kappa = 1.098$ (black solid curve) the curve contains
a segment below $-m_\pi$ (black-dashed curve), which corresponds to a super-critical strength.
}
\label{fig:u-skyrme-mu0}
\end{center}
\end{figure}

In this paper we consider pion systems with $\mu=0$.
Then a nonzero pion density is only possible at $T > 0$.
For each given temperature the particle density is found by self-consistently
solving eq.~(\ref{eq:n}) which now has the following form
\begin{equation}
n\, =\, \frac{g}{2 \pi^2}  \int_0^\infty dk\, k^2\,
\left\{ \exp{\left[\frac{E(\bs k,n)}{T}\right]} - 1\right\}^{-1} \,,
\label{eq:n-mf2}
\end{equation}
where $E(\bs k,n)$ is given in eq.~(\ref{eq:spe}).
From eq.~(\ref{eq:n-mf2}) it is evident that the single-particle energy at
$|\bs k| \to 0$,  $E(\bs 0,n) = m + U(n)$, must be positive or zero.
Otherwise, the occupation numbers become negative.
The potential $U(n)$ is shown in Fig.~\ref{fig:u-skyrme-mu0} for several values
of the parameter $\kappa = A/A_{\rm c}$, characterizing the strength of the
attractive interaction.
Here $A_{\rm c} = 2\sqrt{mB}$ is the critical value of $A$ at which the minimum
of the potential reaches the energy level $-m$.
Below we choose $\kappa$ as a variational parameter to characterize the vicinity
to the critical point.
%

\subsection{Onset of Bose condensation}
The values of $\kappa \ge \kappa_{\rm c} = 1$ lead to a crossing of $-m$
level and appearance of the density interval, where the function $E(\bs 0,n)$
is negative.
The end points of this interval are determined from equation
\begin{equation}
U(n)\, +\, m =\, 0 \,.
\label{eq:condition-condens}
\end{equation}
With $U(n)$ from (\ref{eq:mf1}) the solutions of this equation are
\begin{equation}
n_1 \, =\, \sqrt{\frac{m}{B}} \left( \kappa - \sqrt{\kappa^2 - 1} \right)\,,
\qquad
n_2 \, =\, \sqrt{\frac{m}{B}} \left( \kappa + \sqrt{\kappa^2 - 1} \right) \,.
\label{eq:n1-n2-10}
\end{equation}
In the interval $n_1 < n < n_2$ the integral in eq.~(\ref{eq:n}) is
not positively definite, and therefore such densities can not be realized in an
equilibrium system.
At $\kappa > \kappa_{\rm c}=1$, the change of the pion density
from $n = n_1$ to $n = n_2$ is only possible via the condensation of pions
in zero-momentum mode, $|\bs k| = 0$, so that their total density jumps from
$n = n_1$ to $n = n_2$.
As it is seen from eq.~(\ref{eq:n1-n2-10}) the critical value of parameter
$A$ is obtained when both roots coincide, i.e. when $\kappa = 1$ or
$A=A_{\rm c} = 2\sqrt{m B}$.
For parameter $B$ specified above,
$A_{\rm c} = 2\sqrt{10}m_\pi b \approx 395$~MeV$\cdot$fm$^3$.
The corresponding critical pion density, when the minimum of the potential reaches
the level $-m_{\pi}$, is $n_0 = A_c/2B = \frac{1}{\sqrt{10}b}\approx 0.7$~fm$^{-3}$.
\begin{figure}
\begin{center}
\centering
\includegraphics[width=0.7\textwidth]{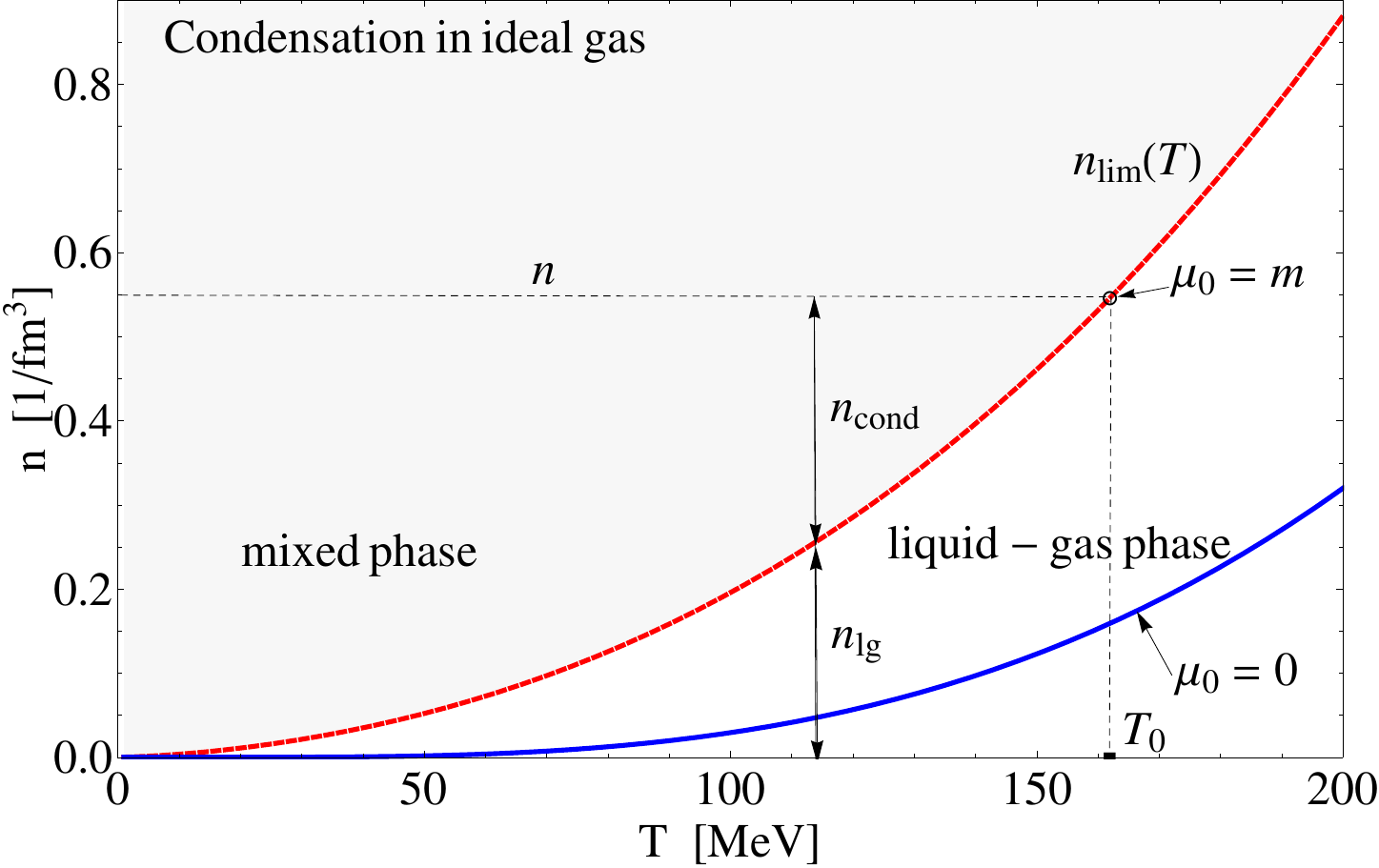}
\centering
\caption{Particle density versus temperature for the noninteracting pion gas
with $\mu_0 = m$ (dashed red curve labeled as $n_{\rm lim}$). This curve
separates the gas phase from the phase with the Bose condensate (dashed area).
The same dependence corresponds to the
interacting pion system with mean field $U(n) = -m$ and $\mu=0$.
For comparison the blue line shows the density of ideal pion gas with $\mu_0=0$.
 }
\label{fig:n-lim}
\end{center}
\end{figure}

The limiting density of Bose particles, $n_{\rm lim}(T)$, just before the formation of
Bose condensate, i.e. at $U(n) = - m$, is the same as in the ideal gas at $\mu_0 = m$
\begin{equation}
n_{\rm lim}(T)\, =\, \frac{g}{2 \pi^2}  \int_0^\infty dk\, k^2\,
\left[ \exp{\left( \frac{\sqrt{m^2 + k^2} - m }{T} \right)} - 1\right]^{-1} \,.
\label{eq:n-lim}
\end{equation}
In Fig.~\ref{fig:n-lim} this dependence is depicted as a red-dashed line, which
separates the normal phase from the phase with the Bose condensate.
We come to the general conclusion: for mean-field potentials deeper than $-m$, the
equilibrated bosonic system will develop a Bose condensate.

\subsection{Energy density and pressure in the liquid-gas phase}
Let us consider first the interacting bosonic system without condensate, i. e. when
$U(n)+m>0$ for all n.
We call this state as a liquid-gas phase, to distinguish it from a
weakly-interacting pion gas.
In the case of $\mu = 0$ the pressure is expressed as
(see (\ref{eq:pressure-int-3}))
\begin{equation}
p(T)\, =\, \frac{g}{6 \pi^2}  \int_0^\infty dk\,\frac{k^4}{\sqrt{m^2 + k^2}}\,
\left\{ \exp{\left[ \frac{\sqrt{m^2 + k^2} + U(n)}{T} \right]}\, -\,
1\right\}^{-1}\, +\, P^{\rm ex}(n) \,,
\label{eq:press}
\end{equation}
where the particle density $n(T)$ is obtained from eq.~(\ref{eq:n-mf2})
assuming $U(n) > -m$.
Here the excess pressure is obtained using the condition of the thermodynamic
consistency (\ref{eq:consistency}) with $U(n)$ from eq.~(\ref{eq:mf1}),
\begin{equation}
P^{\rm ex}(n)\, =\, \int^n_0 dn^\prime\, n'\,\frac{\partial U(n')}{\partial n'}\,
=\, -\,\frac{1}{2} \, A\,n^2\, +\, \frac 23\, B\,n^3 \,.
\label{eq:exess-p}
\end{equation}
%
%
\begin{figure}
\begin{center}
\centering
\includegraphics[width=0.7\textwidth]{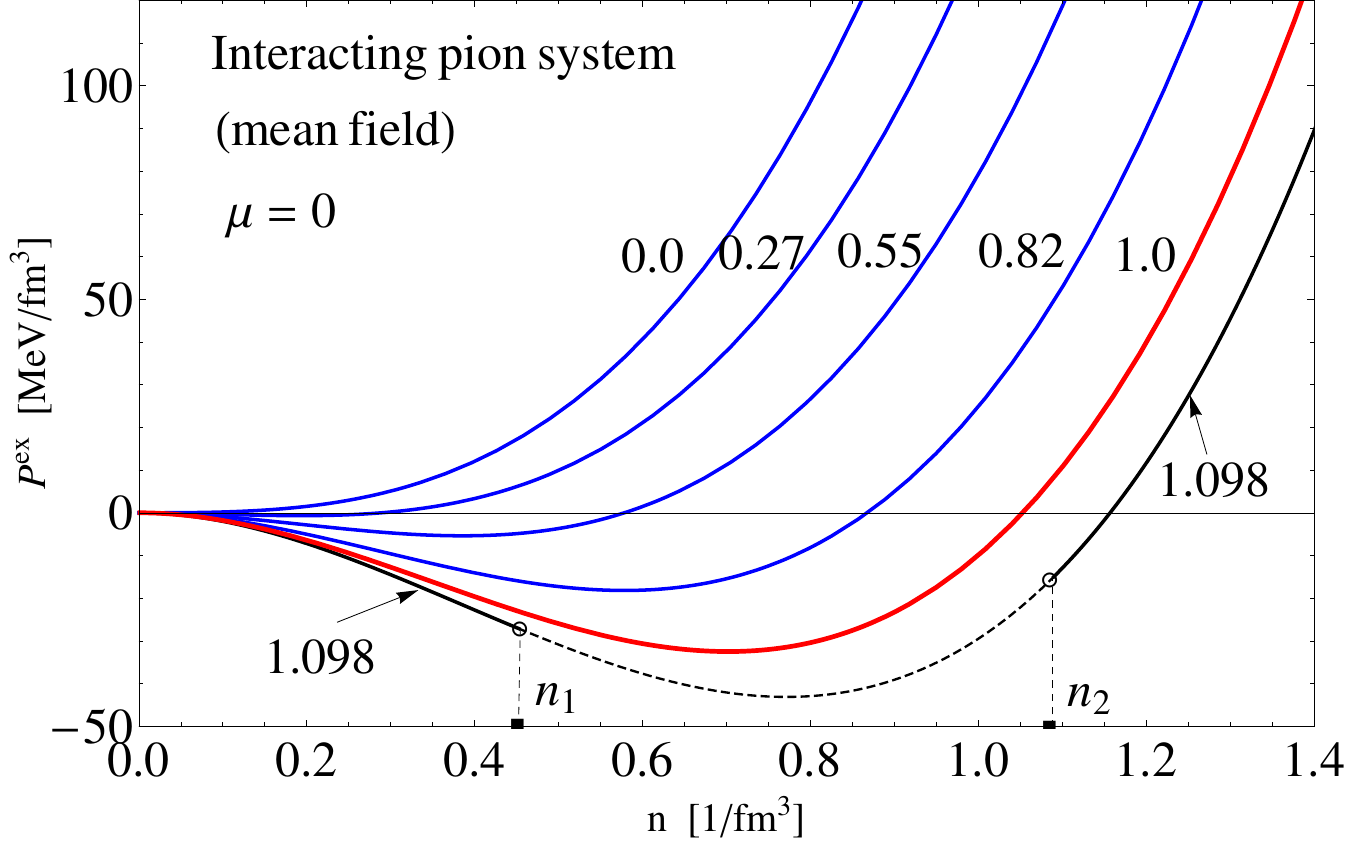}
\centering
\caption{ Excess pressure versus pion density
for interacting pion system with $\mu=0$ for different
values of parameter $\kappa$ indicated in the figure.
The lowest curve calculated for $\kappa = 1.098$ contains a segment $n_1 < n < n_2$
(black dashed line), where $U(n) + m_\pi < 0$.
}
\label{fig:pex-skyrme-mu0}
\end{center}
\end{figure}
Fig.~\ref{fig:pex-skyrme-mu0} shows the excess pressure as a function of the
pion density for several values of $\kappa$.
One can see that it is negative at densities below $3A/4B = \frac 32 \kappa n_0$.

The energy density in the liquid-gas phase is obtained from
eq.~(\ref{eq:energy-den-int}) with $U(n)$ from eq.~(\ref{eq:mf1})
\begin{eqnarray}
\varepsilon(T)\,  =\, g \int \frac{d^3k}{(2\pi )^3}\,\sqrt{m^2 + {\bs k}^2}\,
\left\{ \exp{\left[ \frac{\sqrt{m^2 + \bs k^2} + U(n)}{T} \right]}\, -\,
1\right\}^{-1}\, +\, \varepsilon^{\rm ex}(n) \,.
\label{eq:eden-skyrme5}
\end{eqnarray}
where, in accordance with (\ref{eq:mf1}) and (\ref{eq:exess-p}) the
excess-energy density is
\begin{eqnarray}
\varepsilon^{\rm ex}(n)\,  \equiv \,
n\, U(n)\, -\, P^{\rm ex}(n)\, =\, -\, \frac{1}{2}\,A\,n^2\, +\, \frac 13\, B\,n^3  \,.
\label{eq:ex-eden-skyrme}
\end{eqnarray}
It is interesting to note that the critical condition
(\ref{eq:condition-condens}) corresponds to the requirement that the total
energy density at $T=0$, i.e. $m\cdot n+\varepsilon^{\rm ex}(n)$, has its extremum.
Indeed, the densities $n_1$ and $n_2$ given by eq. (\ref{eq:n1-n2-10})
correspond to the maximum and minimum of this function.
Therefore, it is quite natural that the Bose condensate is formed
at density $n=n_2$.

\section{Self-consistent solution for pion-condensed phase}
\label{sec:condensed-phase}

\subsection{Particle density}
\label{sec:pd}
\begin{figure}
\begin{center}
\centering
\includegraphics[width=0.9\textwidth]{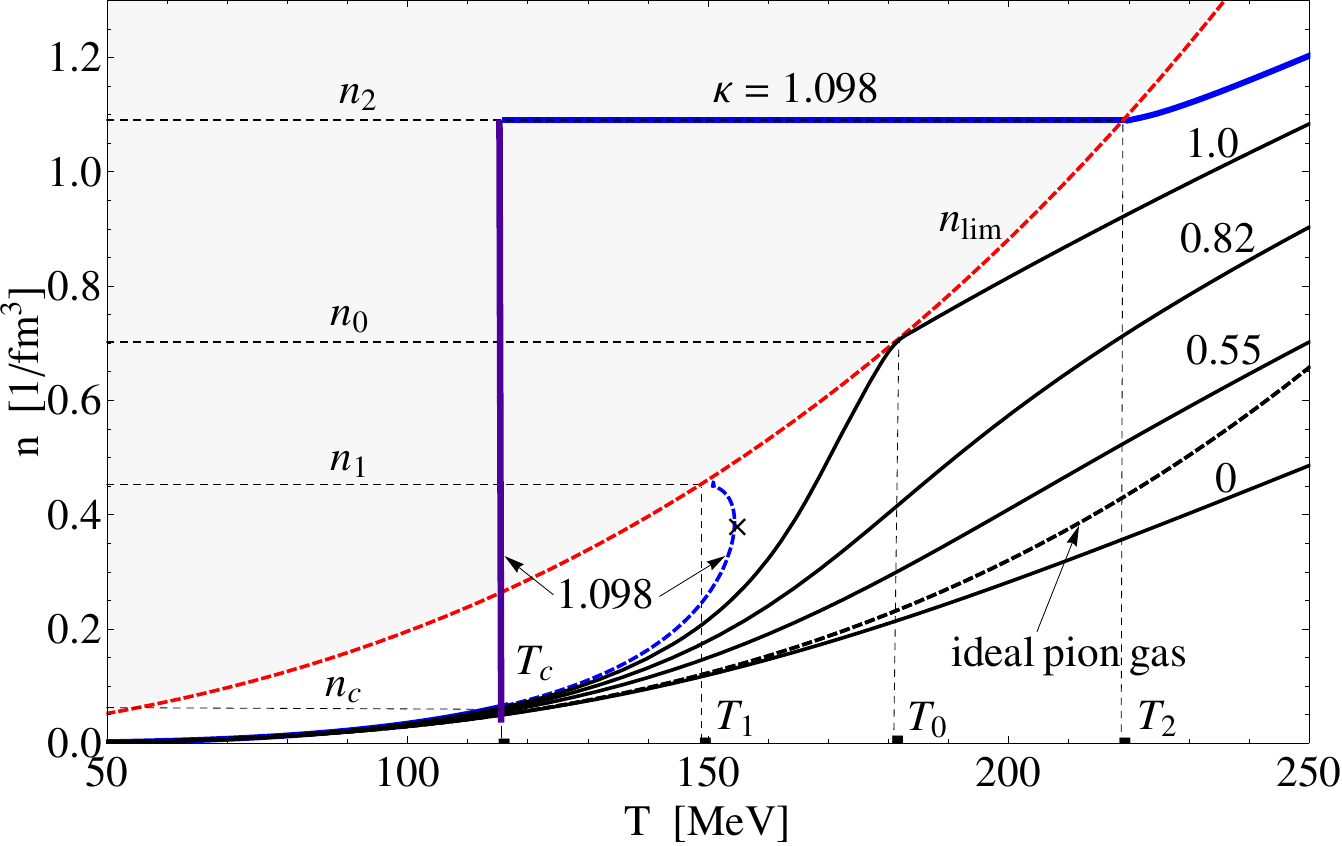}
\centering
\caption{ Particle density versus temperature for interacting pion
system with $\mu = 0$.
The curves are labeled by the values of the interaction parameter
$\kappa = A/A_{\rm c}$.
The densities $n_1,\, n_2$ and $n_0$ are introduced in Fig.~\ref{fig:u-skyrme-mu0}.
The temperatures $T_1,\, T_2$ and $T_0$ indicate the points where these
densities cross the limiting density line $n_{\rm lim}(T)$ (red dashed curve).
For $\kappa = 1.098$ the lower piece of the blue dashed curve corresponds to
the meta-stable states, while  the upper piece is unstable.
The phase transition from liquid-gas phase to mixed phase occurs at
$T_{\rm c} = 116$~MeV and $n_{\rm c}=0.065$~fm$^{-3}$.
For the non-interacting pion system the condensate appears above the line
$n_{\rm lim}$ in the shaded area.
}
\label{fig:tn-depend-skyrme-mu0}
\end{center}
\end{figure}

Let us return to the equation (\ref{eq:n-mf2}), which determines the particle density
as a function of temperature. The solutions $n(T)$ have been found iteratively for several
values of $\kappa$ as shown in Fig.~\ref{fig:tn-depend-skyrme-mu0}.
The limiting density $n_{\rm lim}(T)$, eq.~(\ref{eq:n-lim}), is also shown
in this figure.
One can see that the critical value $\kappa_{\rm c} = 1.0$ separates two
qualitatively different regimes.
At $\kappa < \kappa_{\rm c}$ the curves $n(T)$ are continuous, while at
$\kappa > \kappa_{\rm c}$ they break down in two segments with a gap in between.
One can check that this gap appears exactly between densities $n_1$ and $n_2$,
where $U(n) + m_\pi < 0$, see Fig.~\ref{fig:u-skyrme-mu0}.
For parameter $\kappa = 1.098$ 
the lower branch exhibits a kind of backhanding while approaching the limiting
density $n_{\rm lim}(T)$ from below (the blue-dashed segment)
\footnote{In ref.~\cite{anch-2016} the
appearance of such a bending point (cross in Fig.~\ref{fig:tn-depend-skyrme-mu0})
was interpreted as a ``limiting'' temperature.}.
The appearance of such an unstable branch signals a first order phase transition
in the system, i.e. the Bose condensation of pions.
The critical temperature $T_{\rm c}$ is determined as a crossing point of
pressure curves for the liquid-gas and condensate phases. This is graphically
shown in Fig.~\ref{fig:tp-skyrme-mu0}, which will be discussed in the next subsection.
At $T>T_{\rm c}$ the liquid-gas branch becomes metastable. The barrier between the two
phases finally disappears at the end point marked by the cross (it corresponds to the
banding point in Fig.~\ref{fig:tn-depend-skyrme-mu0}.
Obviously, if we reach the temperature $T_{\rm c}$ and continue to pump energy
into the multi-pion system, it will experience a phase transition leading to
the formation of the Bose condensate, even in the system with $\mu = 0$.
As a consequence the pion density will jump along the line $T = T_{\rm c}$ from
the value $n_{\rm c} = 0.12$~fm$^{-3}$ to the value $n_2 = 1.09$~fm$^{-3}$.
Because of the jump this is certainly a first order phase transition.
It is rather obvious that with further increase of temperature the pion system
will evolve along the horizontal line $n = n_2$ from
$T_{\rm c} = 116.1$~MeV till $T_2 = 219$~MeV, as shown in
Fig.~\ref{fig:tn-depend-skyrme-mu0} (for $\kappa = 1.098$).
\begin{figure}
\begin{center}
\centering
\includegraphics[width=0.7\textwidth]{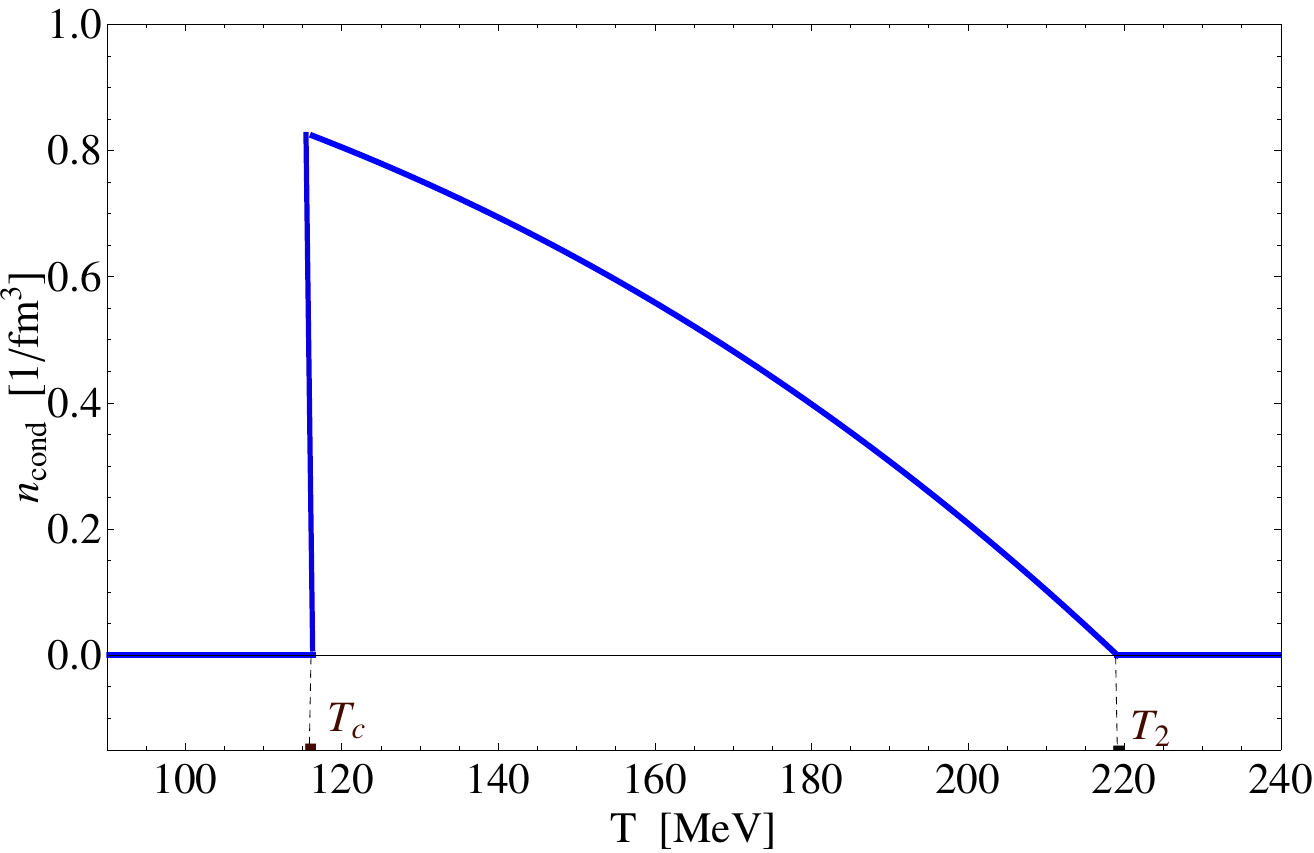}
\centering
\caption{ The density of condensed pions versus temperature for the interacting pion
system with the attraction parameter $\kappa = 1.098$.
The  important temperatures are: $T_{\rm c} = 116.1$~MeV, where the condensate phase
becomes stable, and $T_2=219$~MeV, where the condensate disappears }
\label{fig:tn-condens-skyrme-mu0}
\end{center}
\end{figure}

In a standard treatment of the Bose-Einstein condensation (see for instance
\cite{leggett-2006}), above $T_{\rm c}$ the particle density consists of two
contributions: ``gas-liquid'' particles and ``condensate'' particles.
Therefore, at $T > T_{\rm c}$ we should represent the total particle density
$n$ as
\begin{eqnarray}
n \,  =\, n_{\rm cond}(T)\, +\, \frac{g}{2 \pi^2}  \int_0^\infty dk\, k^2\,
\left[ \exp{\left( \frac{\sqrt{m^2 + k^2} - m }{T}\right)} - 1\right]^{-1} \,,
\label{eq:n-sce-cond}   
\end{eqnarray}
where $n_{\rm cond}$ is the density of condensed pions.
This equation is valid in our specific case too, where the evolution of the
system goes along the constant density line $n = n_2$.
Indeed, for every temperature $T$ from the interval $T_{\rm c} < T \le T_2$
(see Fig.~\ref{fig:tn-depend-skyrme-mu0})
the density of particles is $n_2 = n_{\rm lim}(T) + n_{\rm cond}(T)$ and the
value of the mean field is $U(n_2) = - m$.
The behavior of $n_{\rm cond}(T)$ is shown in Fig.~\ref{fig:tn-condens-skyrme-mu0}.
Hence, one should consider eq.~(\ref{eq:n-sce-cond}) as a selfconsistent
description of the pion condensate in the framework of the mean-field approach.
%

\subsection{Pressure}
\label{sec:pressure}
%
In the mixed phase with the Bose condensate the total particle density is fixed
at $n = n_2$, and thus the pressure can be expressed as
\begin{equation}
p_{\rm cond}(T)\, =\, g\int \frac{d^3k}{(2\pi)^3}\,
\frac{{\bs k}^2}{\sqrt{m^2 + {\bs k}^2} }\,
\left\{ \exp{\left[ \frac{\sqrt{m^2 + \bs k^2} - m}{T} \right]}\, -\,
1\right\}^{-1}\, +\, P^{\rm ex}\big(n_2\big) \,,
\label{eq:pressure-condens}
\end{equation}
where $P^{\rm ex}(n_2)$ is given by eq.~(\ref{eq:exess-p}) with $n = n_2$.
\begin{figure}
\begin{center}
\centering
\includegraphics[width=0.7\textwidth]{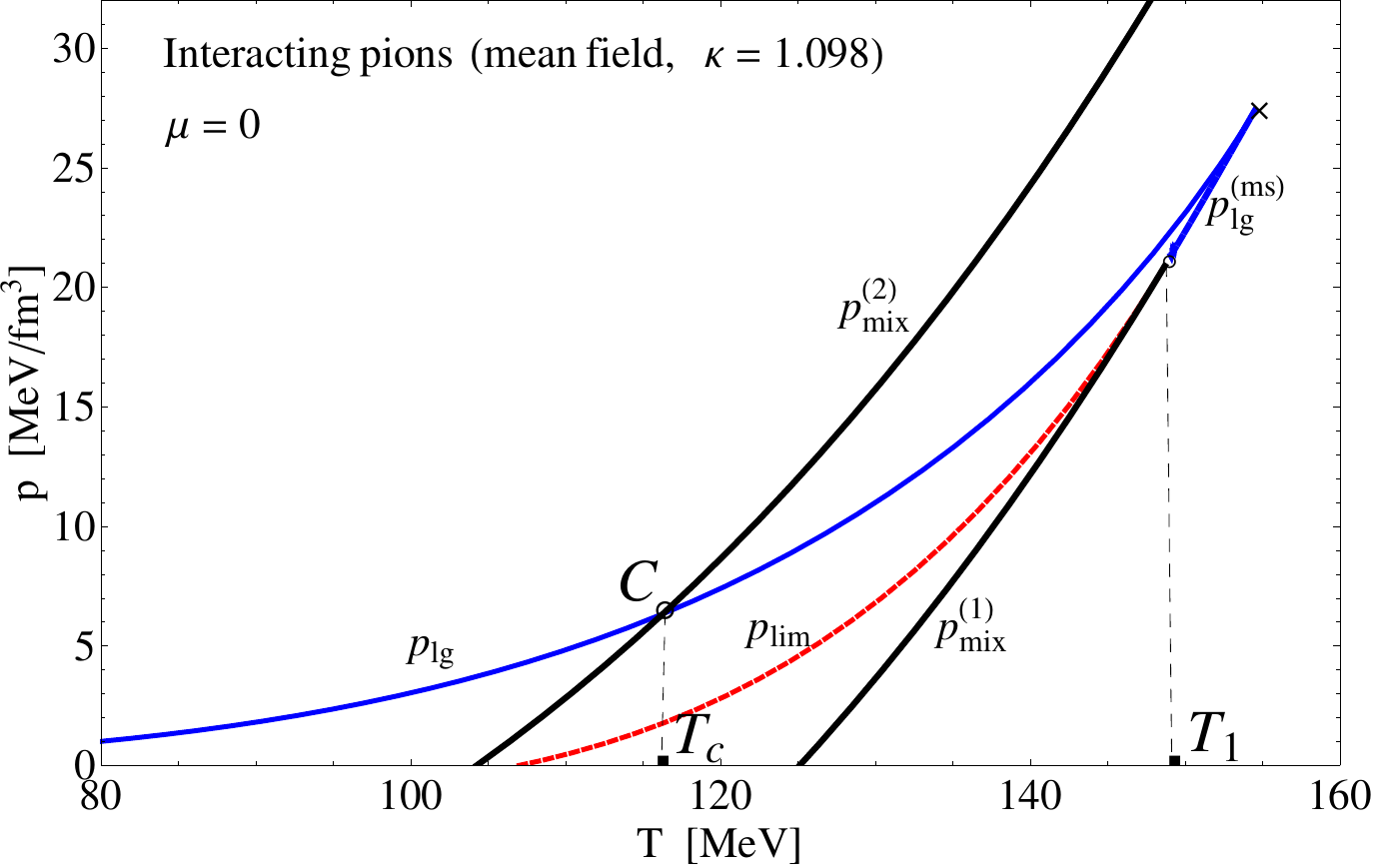}
\centering
\caption{ Pressure versus temperature for an interacting pion system
with $\mu=0$ for super-critical interaction parameter $\kappa = 1.098$.
Solid blue line labeled as $p_{\rm lg}$ corresponds to the pressure in the liquid-gas
phase, the dark blue line labeled as $p_{\rm lg}^{\rm (ms)}$ corresponds to
meta-stable states in the liquid-gas phase,
the ``cross'' marks the turning-point.
Solid black lines labeled as $p_{\rm mix}^{(1)}$ and $p_{\rm mix}^{(2)}$
correspond to the pressures in the mixed phase (condensate plus liquid-gas)
for the pion densities $n = n_1$ and $n = n_2$, respectively.
C is the critical point.
The branch $p_{\rm mix}^{(1)}$ is unstable.
}
\label{fig:tp-skyrme-mu0}
\end{center}
\end{figure}
One should bear in mind that the condensate particles with $k = 0$ give no
contribution to the kinetic part of the pressure (first term), but contribute to
the interaction pressure via $P^{\rm ex}$ (second term).
Note, that at $\kappa = 1.098$ $P^{\rm ex}$ is negative for pion densities in
the interval $0 < n < 1.16$~fm$^{-3}$, see Fig.~\ref{fig:pex-skyrme-mu0}.
Figure \ref{fig:tp-skyrme-mu0} shows several branches of pressure
representing different phases.
The critical temperature, $T_{\rm c} = 116$~MeV, is obtained as the crossing
point of two branches, representing the liquid-gas pressure for
$\kappa = 1.098$ and pressure of the mixed phase at $n=n_2$.
This explains why the Bose condensate appears only above $T_{\rm c}$, when the
additional thermal pressure compensates the negative contribution of $P^{\rm ex}$.
At $T < T_{\rm c}$ the mixed-phase branch is metastable.

\subsection{Energy density}
\label{sec:ed}
In the mixed phase the energy density
consists of the kinetic part, $\varepsilon_{\rm kin}(T)$, which is produced by
particles in the liquid-gas phase with density $n_{\rm lg}(T) = n_{\rm lim}(T)$,
and the condensate particles with density $n_{\rm cond}(T)$, due to the
particles, with zero momentum.
In accordance with self-consistent solution of eq.~(\ref{eq:n-sce-cond}),
in the mixed phase, a sum of these densities remains constant,
$n_{\rm lg}(T) + n_{\rm cond}(T) = n_2$.
This constant density $n_2$ determines the excess pressure
$P^{\rm ex}(n_2)$ and excess energy density $\varepsilon^{\rm ex}(n_2)$
in the mixed phase.
\begin{figure}
\begin{center}
\centering
\includegraphics[width=0.8\textwidth]{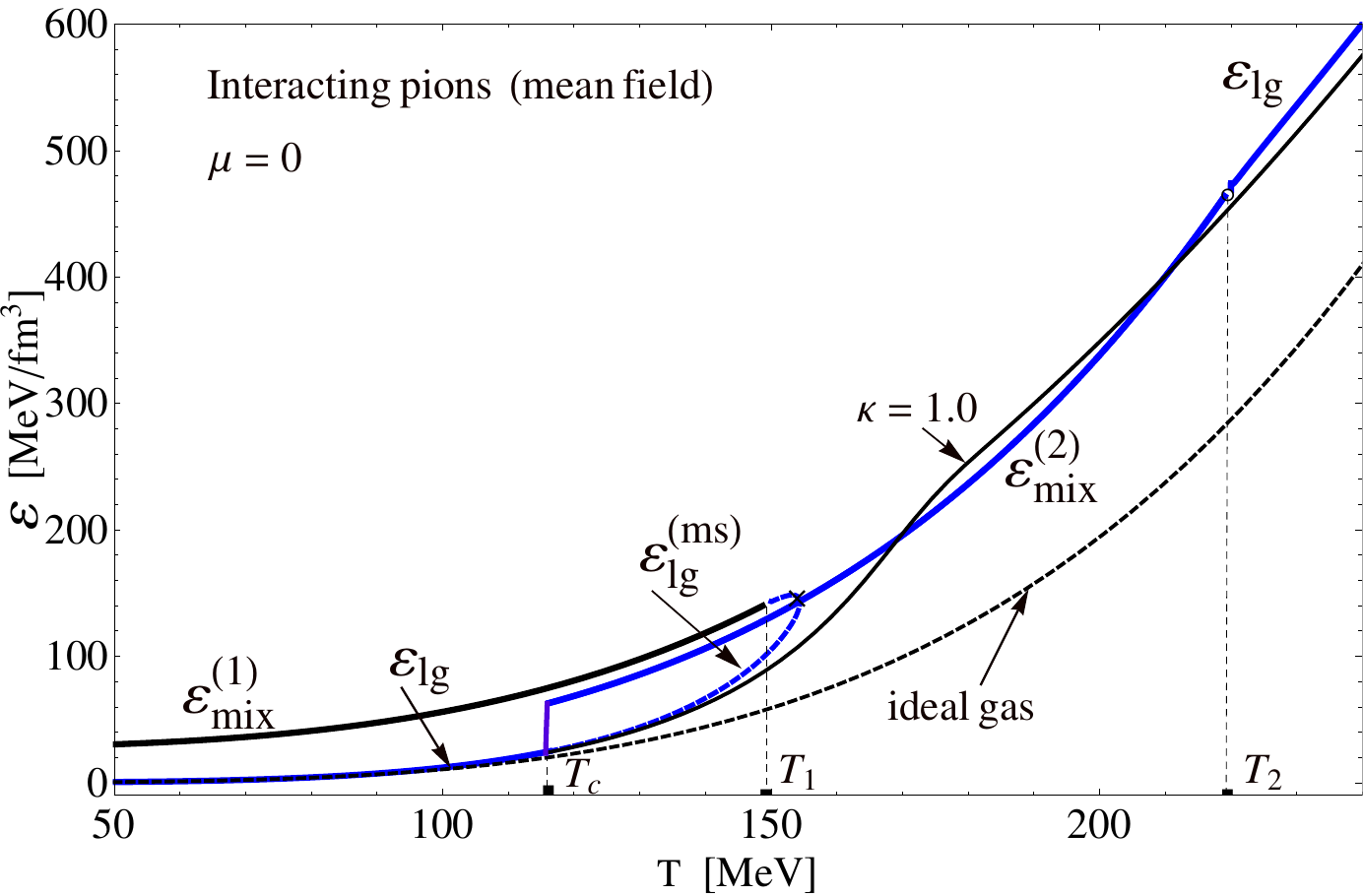}
\centering
\caption{ The energy density versus temperature for an interacting pion system with $\mu=0$.
The black solid curve labeled as $\kappa = 1.0$ corresponds to the energy density for the
parameter $\kappa = \kappa_{\rm c}$. The
blue solid curve, which consists of several segments, labeled as
$\varepsilon_{\rm lg}$ and $\varepsilon_{\rm mix}^{(2)}$, corresponds to $\kappa = 1.098$.
The latter corresponds to the mixed phase with constant particle density $n=n_2$ in
the temperature interval $T_{\rm c} \le T \le T_2$.
The energy density of meta-stable states in liquid-gas phase for $\kappa = 1.098$ is
labeled as $\varepsilon_{\rm lg}^{\rm (ms)}$ (blue dashed segment).
The energy density in the mixed phase $\varepsilon_{\rm mix}^{(1)}$ calculated
for the constant density $n=n_1$ is depicted as black solid segment.
The energy density of metastable states in the liquid-gas phase for $\kappa = 1.098$ is
labeled as $\varepsilon_{\rm lg}^{\rm (ms)}$ (blue dashed segment).
The ideal gas energy density is depicted as black dashed curve.
}
\label{fig:energy-density-mu0}
\end{center}
\end{figure}
Therefore, the energy density in the mixed phase at $n = n_2$ reads
\begin{eqnarray}
\varepsilon_{\rm mix}^{(2)} = m n_{\rm cond}(T) +
g\! \int_{|\bs k| \neq 0} \frac{d^3k}{(2\pi )^3} \sqrt{m^2\! +\! {\bs k}^2}
\left[ \exp{\! \left( \frac{\sqrt{m^2 + \bs k^2} - m}{T} \right)} - 1\right]^{-1}
\! + \varepsilon^{\rm ex}(n_2) ,
\label{eq:eden-skyrme-cond}
\end{eqnarray}
where the condition $U(n_2) = - m$ has been used.
Using eq.~(\ref{eq:ex-eden-skyrme}) and the expression for $n_{\rm cond}(T)$
from eq.~(\ref{eq:n-sce-cond}), one can rewrite this expression as
\begin{equation}
\varepsilon_{\rm mix}^{(2)}\, =\,
g \int_{|\bs k| \neq 0} \frac{d^3k}{(2\pi )^3} \left(\sqrt{m^2 + {\bs k}^2} - m \right)
\left[ \exp{\left( \frac{\sqrt{m^2 + \bs k^2} - m}{T} \right)} - 1\right]^{-1}
 - P^{\rm ex}(n_2) \,.
\label{eq:eden-skyrme-cond2}
\end{equation}
In Fig.~\ref{fig:energy-density-mu0} we present the energy densities for different
systems: interacting pions with $\kappa = 1.098$ (blue solid curve), interacting
pions at critical value $\kappa=\kappa_{\rm c} = 1.0$ (black solid curve)
and the ideal pion gas (black dashed curve).
One can see that for $\kappa = 1.098$ the model predicts upward jump
of the energy density of about $30$~MeV/fm$^3$ (latent heat), at critical temperature
$T_{\rm c} = 116.1$~MeV.
This is another manifestation of the first order phase transition.
One can see in Fig.~\ref{fig:energy-density-mu0} that in a wide
interval of temperatures the energy density of the critical system with
$\kappa = 1$ is larger than that in the condensed phase with $\kappa = 1.098$
represented by $\varepsilon^{(2)}_{\rm cond}$.

\section{Concluding remarks}
\label{sec:conclusions}
In this paper we have presented a thermodynamically consistent method to
describe dense bosonic systems at high temperatures and zero chemical potential.
A central step of this approach is to solve the self-consistent equations
(\ref{eq:n-mf2}) or (\ref{eq:n-sce-cond}) for the pion density at a given
temperature.
The crucial point is calculation of the kinetic integral (\ref{eq:n-lim}),
which determines the pion density in the presence of the mean-field
$U(n)$ by eq.~(\ref{eq:mf1}).
This integral is positive definite only if the condition $U(n) \ge - m$
is fulfilled.
If the attractive mean field is so strong that this condition is violated,
the multi-boson system develops a Bose condensate.
Our analysis leads to the conclusion that in presence of the condensate, the
allowed states of the system must satisfy the condition $U(n) + m = 0$,
where $n$ is the total particle density including the condensate.
This very unusual behavior is only possible if the attractive interaction
between bosons is strong enough.
However, as was already mentioned in Sec.~\ref{sec:interacting-ps}, the
empirical data and theoretical calculations, see ref.~\cite{gasser-1989},
show that the pion-pion interaction is rather weak at energies $\le 100$~MeV.
Nevertheless, an additional contribution to the pion mean field can
be provided by
the attractive pion-nucleon interaction in cold nuclear matter, as demonstrated in
refs.~\cite{migdal-1972,migdal-1974,migdal-1978,saperstein-1990},
or by $\rho$-mesons and  baryon-antibaryon pairs at high temperatures, as
considered in refs. \cite{shuryak-1991,Theis}.
Another interesting possibility, studied in refs.
~\cite{mishustin-greiner-1993,son-2001,kogut-2001,toublan-2001,mammarella-2015,
carignano-2017},
is Bose condensation in a pure pionic system with non-zero isospin chemical
potential.
We are planning to consider such interacting systems in the future.

Finally, we would like to point out that on the first site it seems that
the mesonic degrees of freedom may not be appropriate at high particle densities
($n \approx 1\,$fm$^{-3}$) and high temperatures ($T \ge 150\,$MeV)
as considered in this paper.
Meanwhile, the temperature of condensation, $T_{\rm c} \approx 116\,$MeV, lies
in a well determined hadronic sector, hence the melting of pions into quarks
and antiquarks during heating of the system can go directly from the dense phase
of condensed particles.
And vise versa, during cooling of the system soft pions can form a condensate
on the stage of hadronization.

On the other hand, the multi-quark-antiquark
systems can be studied by using either lattice methods, as
in refs.~\cite{brandt-2016,brandt-2017}, or QCD motivated effective models.
For example, in ref.~\cite{mishus-1999} the Nambu-Jona-Lasinio model has been
used to describe baryon-free matter made of quarks and antiquarks, so called
``meso-matter''.
It was found that such systems are characterized by strong attractive and reduced
repulsive interactions, leading to a strong liquid-gas phase transition.
It would be also interesting to investigate the possibility of $q-\bar{q}$
pairing and the Bose-Einstein condensation in such systems.

\section*{Acknowledgements}
D.~A. sincerely thanks the staff of the FIAS, where the major part of this work
was done, for collaboration and warm hospitality.
The work of D.~A. was supported by the Program
"The structure and dynamics of statistical and quantum-field systems"
of the Department of Physics and Astronomy of NAS of Ukraine and
Ukraine-Hungary project "Kinetic and critical phenomenon in nonequilibrium
quantum systems in finite space-time regions".
I.~M. acknowledges financial support from Helmholtz International Center for
FAIR, Germany.
H.S. appreciates the support from J.M. Eisenberg laureatus chair.
The authors thank L.M. Satarov for fruitful discussions.


\end{document}